\documentclass[a4paper,10pt]{llncs}
\usepackage{amssymb}
\usepackage{amsmath}
\usepackage{graphicx}
\usepackage[colorlinks=true,linkcolor=blue,anchorcolor=black,citecolor=green,filecolor=red,menucolor=black,urlcolor=blue,breaklinks=true,pdfhighlight=/P,pdfmenubar=true,pdftoolbar=true,pdfpagelabels=true,pdfstartpage=1,pdfstartview=FitV,pdftitle={Titel},pdfsubject={Thema},pdfauthor={Autor},pdfcreator={Ersteller},pdfproducer={Produzent},pdfkeywords={Stichworte}]{hyperref}	%Macht aus Referenzen interne und externe Links
\usepackage[numbers,sort&compress]{natbib}	%Sortiert Zitate und macht aus [10,11,12,13,14] [10-14]

\urldef{\mailms}\path|{schultz,fricke}@ifl.tu-dresden.de|
\urldef{\mailtk}\path|tobias.kretz@ptv.de|
\newcommand{\keywords}[1]{\par\addvspace\baselineskip
\noindent\keywordname\enspace\ignorespaces#1}

\begin{document}
\mainmatter

\title{Solving the Direction Field for Discrete Agent Motion}

\titlerunning{Directions}

\author{Michael Schultz\inst{1}, Tobias Kretz\inst{2}, and Hartmut Fricke\inst{1}}

\institute{Institute of Logistics and Aviation, Technische Universit\"at Dresden,\\
Hettner Str. 1-3, D-01062 Dresden, Germany\\
\mailms\\
% \url{http://www.ifl.tu-dresden.de}
\and
PTV Planung Transport Verkehr AG,\\
Stumpfstr. 1, D-76131 Karlsruhe, Germany\\
\mailtk
}

\authorrunning{}

\maketitle

\begin{abstract}
Models for pedestrian dynamics are often based on microscopic approaches allowing for individual agent navigation. To reach a given destination, the agent has to consider environmental obstacles. We propose a direction field calculated on a regular grid with a Moore neighborhood, where obstacles are represented by occupied cells. Our developed algorithm exactly reproduces the shortest path with regard to the Euclidean metric.

\end{abstract}

\section{Distances and Directions}

To give robots or simulated pedestrians (agents) their main direction of movement, for at least three decades there have been two main methods: one uses a set of navigation points to steer the agent around obstacles \cite{deBerg1997}, the other one -- which this contribution deals with -- relies on a grid in which each grid cell holds the information on the walking distance to the destination and/or the direction to move on to be on the shortest or roughly quickest path considering the location of obstacles. This grid is often called a ``potential'' or a ``distance look up table''.
An early usage of the notion and method of a potential was made by Khatib \cite{Khatib1986} in robot motion planning. But this potential did not consider the location of obstacles but just held the bee-line distances from the location of the robot to the destination. This makes sense in the motion planning of \emph{autonomous} robots, which only have a very limited knowledge of their environment and need an elaborate method to escape dead-ends anyway \cite{Latombe1991}.

The most prominent and most widely used method that calculates Euclidean distances as a numerical solution of the Eikonal equation \cite{Bruns1895,Frank1927} comparatively fast and with a comparatively small error is the Fast Marching Method (FMM) \cite{Osher1988,Kimmel1998,_Sethian1999}. Concerning computation time the FMM shows optimal worst-case behavior and the relative error in general decreases with increasing distance from the destination, implying that with a finer grid the error can be reduced. With the change of computing power progress from improving single CPU processing to multi-core computation, slight changes in the algorithm of the up-winding scheme have become useful and recently new methods as for example the Fast Sweeping \cite{Zhao2005} and the Fast Iterative Method \cite{Jeong2007} (FSM and FIM) have been introduced. For a study of computation times of such methods see \cite{Gremaud2006} for example.

Apart from optimizing the numerical Eikonal equation solver algorithmically, it's also possible to trade exactness for computation speed and use simpler methods for the calculation. The two most prominent examples for this are a simple flood fill over common edges or common edges and common corners which lead to metrics in vector norms (\ref{eq:dist_allg}), $p=1$ (Manhattan metric) or $p\rightarrow\infty$ (Chebychev metric) respectively. 

\begin{equation}
	d^n_p = \left\|x\right\|_p := \left( \sum^n_{i=1} \left|x_i\right|^p \right) ^{\frac{1}{p}}
	\label{eq:dist_allg}
\end{equation}

With Euclidean metric (i.e. vector norm $p=2$) as correct solution, these methods lead to relative errors which remain constant over distance or even increase, which means that a finer grid size does not improve the precision arbitrarily. However, it is possible to reduce the error by making some slight modifications upon the simple flood fill methods \cite{Kretz2010a}. Just for completeness we want to add that to our knowledge no other than ray tracing methods exist, that are able to reach in general cases the minimal error possible, which is given by the grid resolution \cite{Kretz2010a}. However, with such methods compared to FMM, FSM or FIM normally one pays with a tremendous increase of computation time for a small gain in exactness.

In many models of pedestrian (or robot) dynamics only the desired walking direction (i.e. the gradient of walking distances) but not the walking distances themselves are inputs for the calculation of the dynamics. Therefore in this contribution we put forward a new method to calculate the (in terms of the Euclidean metric) exact walking directions without having to calculate the exact Euclidean walking distances. The method does neither rely on a computation time expensive sorting of distances of currently \emph{active} cells, nor does it even has the need for the calculation of rather computation-time expensive functions (e.g. square roots). This is achieved on the expense that the method can only be used to calculate directions based on shortest distance and not shortest time \cite{Hughes2002,Hoogendoorn2004,Treuille2006,Shopf2008,Kretz2009a,Bleiweiss2009,Steffen2009,Kretz2010c} and that only the directions but not the distances are calculated exactly regarding to the Euclidean metric.

\section{Algorithm}

A common practice for creating a distance potential is a flood fill approach. The following \emph{flood fill} algorithm describes the essential steps for the creation of the potential field (\emph{breadth first search algorithm} applied for von Neumann neighborhood \cite{Lee1961}).

\begin{itemize}
	\item The target cell $T$ is initialized with the distance $d_T=0$, the other $n$ cells get the distance $d_n=\infty$.
	\item Put $T$ in a first-in, first-out queue $Q$.
	\item While cells available in $Q$ do:\\
				$\rightarrow\ \hspace{10mm}$ poll (get and remove) the first cell ($C$) from $Q$\\
				$\rightarrow\ \hspace{10mm}$ set $C$ as center cell\\
				$\rightarrow\ \hspace{10mm}$ for all adjacent cells $C_n$:\\
				$\rightarrow\	\hspace{20mm}$ calculate distance to the target cell regarding to the center\\
				$\null				\hspace{25mm}$ cell $d_n = \min( d_n, d_C + \Delta d)$\\
				$\rightarrow\ \hspace{20mm}$ if $d_n$ changes, put $C_n$ in $Q$
\end{itemize}

The distance between adjacent cells depends on the relative position (diagonal or horizontal/vertical). Using a geometry $G$, non-accessible cells (e.g. environmental obstacles) possess a distances of $\Delta d =\infty$. For $\Delta d$ three different cases exist:
\begin{itemize}
	\item[$\infty$], if $C_n$ is a infrastructure cell,
	\item[$\sqrt{2}$], if $C_n$ is diagonal located regarding to $C$, and
	\item[$1$] , if $C_n$ is horizontal/vertical located regarding to $C$.
\end{itemize}

\begin{figure}
	\centering
	\includegraphics{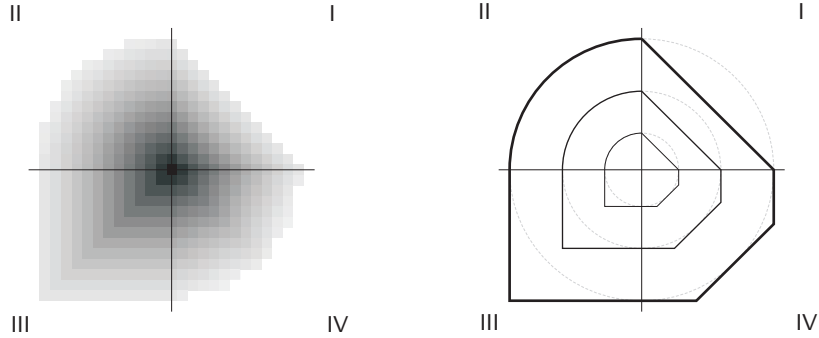}
	\caption{Distance metric characteristics: transition from black to white corresponds with decreasing potential (left), equipotential lines (right).}
	\label{fig:r_range}
\end{figure}

In comparison to the Manhattan (quadrant I, see fig. \ref{fig:r_range}), Euclidean (II) and Chebychev (III) metrics, the flood fill approach for Moore neighborhood points out a fourth distance metric (IV):

\begin{equation}
	\label{eq:dist_man_moore}
	\left\|x\right\| : = \left| \Delta x_i \right| \, + \, \sqrt{2} \ \min( |x_i| ) \; .
\end{equation}

As shown in fig. \ref{fig:r_range} the presented flood fill metric approximates the Euclidean metric in a roughly way and it tends to overestimate the distance (except at the grid symmetry axes). However, the flood fill approach is often used to create the potential field for agent navigation. The corresponding agent rule is: "Choose the cell with the closest distance to the target".
 
\section{Direction Field}

\begin{figure}
	\centering
	\includegraphics{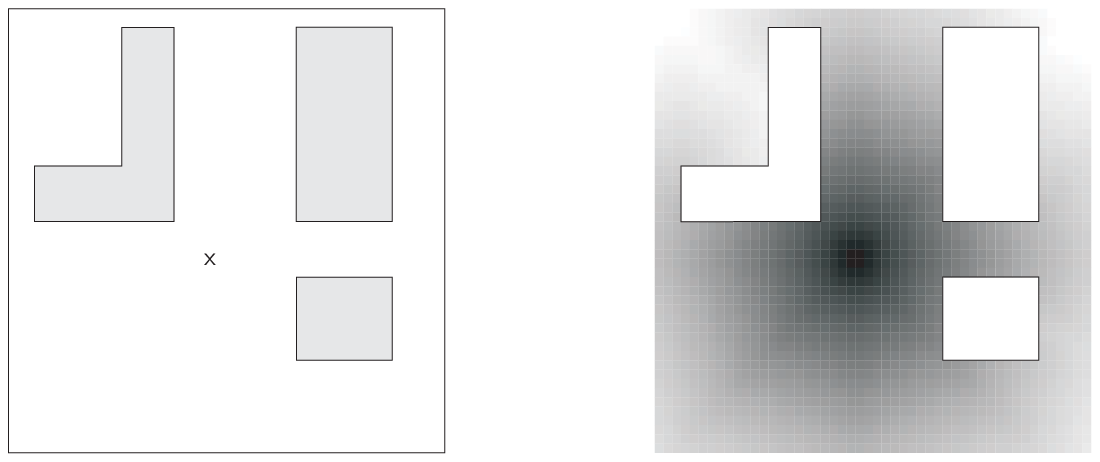}
	\caption{Test scenario with a centered target and three obstacles (left) together with the corresponding color-coded distance field (decreasing target distance from white to black, right).}
	\label{fig:test_scenario}
\end{figure}

Despite to the significant deficiency of the flood fill metric, we will demonstrate that this approach can be used for creating a precise direction field \cite{Schultz2010}. To create a simple direction field $D$ (represents an array of motion vectors) the distance field is created for a test scenario (see fig. \ref{fig:test_scenario}). Close inspections of the color-coded distance field at fig. \ref{fig:test_scenario} (right) already reveals the distance metric for the Moore neighborhood.

\begin{figure}
	\centering
	\includegraphics{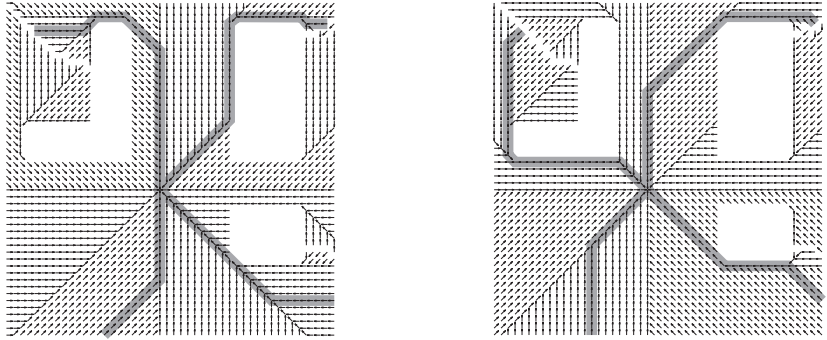}
	\caption{Different characteristic of simple direction field based on the expansion of the flood fill algorithm. Expanding cells clockwise ($D^+$, left) and counter-clockwise ($D^-$, right).}
	\label{fig:r_plus_and_r_minus}
\end{figure}

The flood fill algorithm stores the shortest target distance for each cell, whereas the cell specific distance always will base on a diagonal or horizontal/vertical located adjacent center cell. The location of the upstream center cell depends on the processing sequence of the adjacent cells, which is defined by the algorithm. To determine the direction field, the derived cell based motion vector points to the particular center cell. Detailed verifications show, that the characteristics of the simple direction field depends on the particular implementation (see fig. \ref{fig:r_plus_and_r_minus}). First implementations determine the sequence of the adjacent randomly, followed by sequences where the cells are clockwise and counter-clockwise calculated. 

As fig. \ref{fig:r_plus_and_r_minus} shows, characteristic Moore neighborhood patterns evolve and constantly alternate in steps of $\frac{\pi}{4}$. The change of the sequence form clockwise to counter-clockwise results in a complementing structure ($D^+\rightarrow D^-$, fig. \ref{fig:r_plus_and_r_minus}). Areas those before contain diagonal vectors are now contain horizontal/vertical motion direction vectors. Considering the Moore metric (\ref{eq:dist_man_moore}) the designated paths are equivalent to their walking distance.

\begin{figure}
	\centering
	\includegraphics{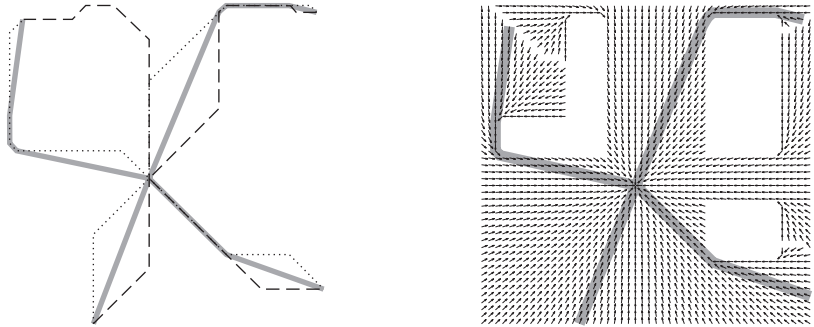}
	\caption{Combining the particular motion direction vectors from $D^+$ and $D^-$ results in the precise direction field.}
	\label{fig:r_derived}
\end{figure}

Each of the previously created direction fields ($D^+$ and $D^-$) represents one component of the final direction field. At the first step the cell based motion direction vectors have to be combined, so sequently aligned vectors are summed up to the point where the direction changes (fig. \ref{fig:r_derived}, left). Now each particular cell contains a diagonal and a horizontal/vertical direction component (fig. \ref{fig:r_derived}, right). The final direction field indicates no directional artifacts and the declared paths are consistent to the Euclidean metric. 

\section{Summary and Outlook}

The proposed algorithm efficiently prevents the directional artifacts by combining two different simple direction fields, which are based on Moore metric distance calculation. Due to the fact, that the flood fill algorithm for creating a distance potential is widely used in agent simulation, our enhanced approach provides a fundamental contribution to existing simulation systems. The introduced direction field is an essential part of the route planning component inside the virtual terminal environment \cite{Schultz2010} of the Institute of Logistics and Aviation. Future investigations regarding to group dynamic behavior or route planning in the airport terminal environment at normal operations or emergency cases will benefit from our proposed algorithm.

%\nocite{_PED2001,_TGF2001,_ACRI2002,_ACRI2006}
%\bibliographystyle{unsrt}
\bibliographystyle{utphys_quotecomma}
\bibliography{Directions_ACRI10}

\end{document}